\documentclass[letterpaper, 10 pt, conference]{ieeeconf}  % Comment this line out
\IEEEoverridecommandlockouts                              % This command is only
\title{\LARGE \bf Using Virtual Humans to Understand Real Ones}

\author{Katie Hoemann$^{1}$, Behnaz Rezaei$^{2}$\\
Project Advisors: Prof. Stacy C. Marsella $^{3}$, Prof. Sarah Ostadabbas$^{2}$ % <-this % stops a space
\thanks{$^{1}$Katie Hoemann is with the Interdisciplinary Affective Science Laboratory, Department of Psychology, Northeastern University, MA, USA.}
\thanks{$^{2}$Behnaz Rezaei and Sarah Ostadabbas are with the Augmented Cognition laboratory, Electrical and Computer Engineering Department, Northeastern University, MA, USA (corresponding author's e-mail: ostadabbas@ece.neu.edu).}
\thanks{$^{3}$Stacy C. Marsella is with the College of Computer and Information Science and Department of Psychology, Northeastern University, MA, USA.}
}

\usepackage{array}
\usepackage{ifpdf}
\usepackage{graphicx}
\usepackage{epsfig} % for postscript graphics files

\usepackage[cmex10]{amsmath}
\usepackage{amsfonts}
\usepackage{algorithmic}
\usepackage{algorithm2e}
\usepackage{mdwmath}
\usepackage{colortbl}
\usepackage{hhline}
\usepackage{multirow}
\usepackage[caption=false,font=footnotesize]{subfig}
\usepackage{url}
\usepackage{fixltx2e}
\usepackage{float}
\usepackage{balance}
\usepackage[english]{babel}
\usepackage{comment}

\newcommand{\eqnref}[1]{Equation~(\ref{eqn:#1})}
\newcommand{\figref}[1]{Fig.~\ref{fig:#1}}
\newcommand{\tblref}[1]{Table~\ref{tbl:#1}}

\newcommand{\figvignet}{
\begin{figure}
  \centering
  \includegraphics[width=0.48\textwidth,  trim=0in 0in 0in 0in, clip=true]{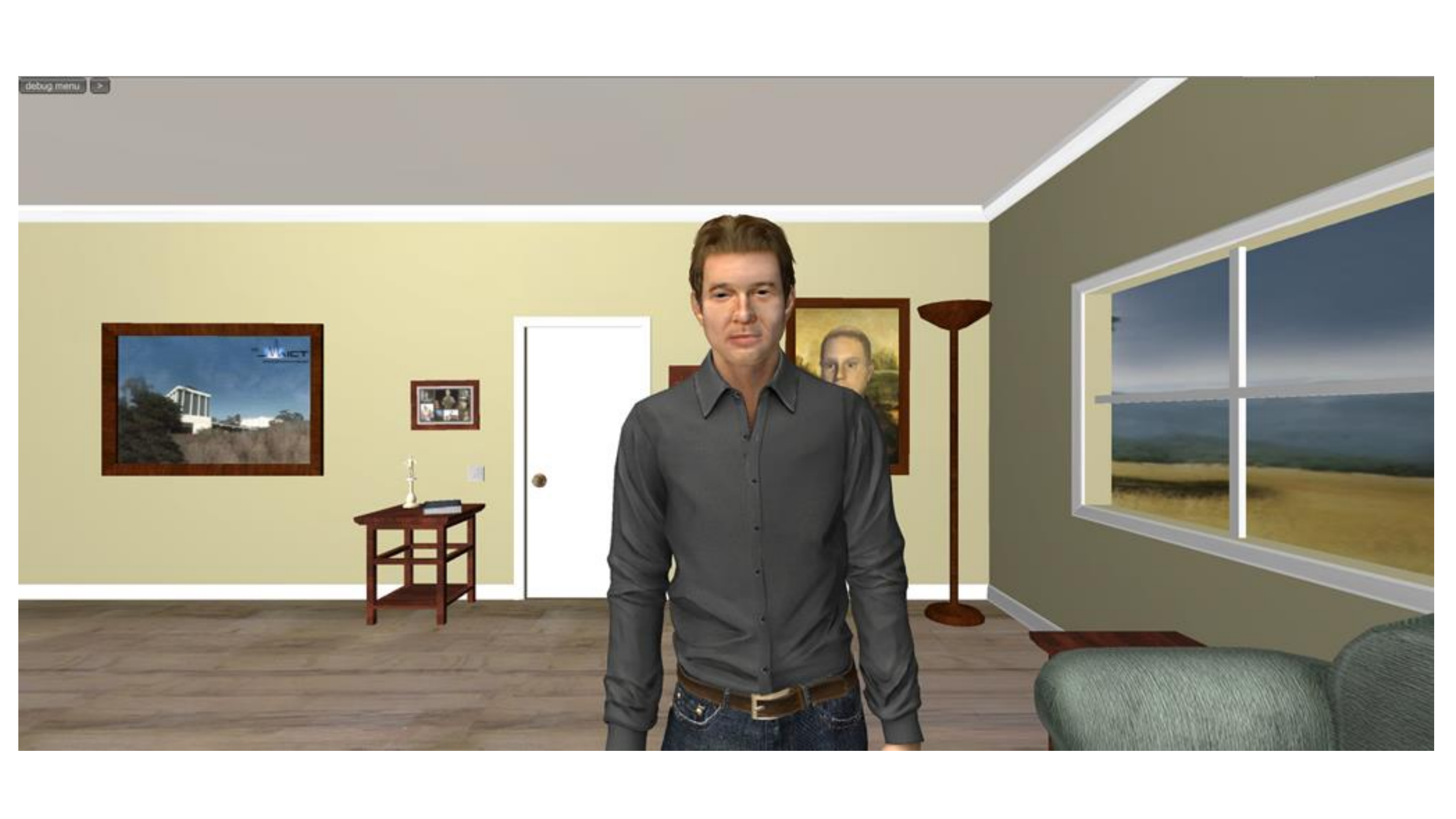}
  \caption{Scenario 5, house scene, with both gaze and gesture on }
\label{fig:vignet}
\end{figure}
}

\newcommand{\figpleasure}{
\begin{figure}
  \centering
  \includegraphics[width=0.48\textwidth,  trim=0in 0in 0in 0in, clip=true]{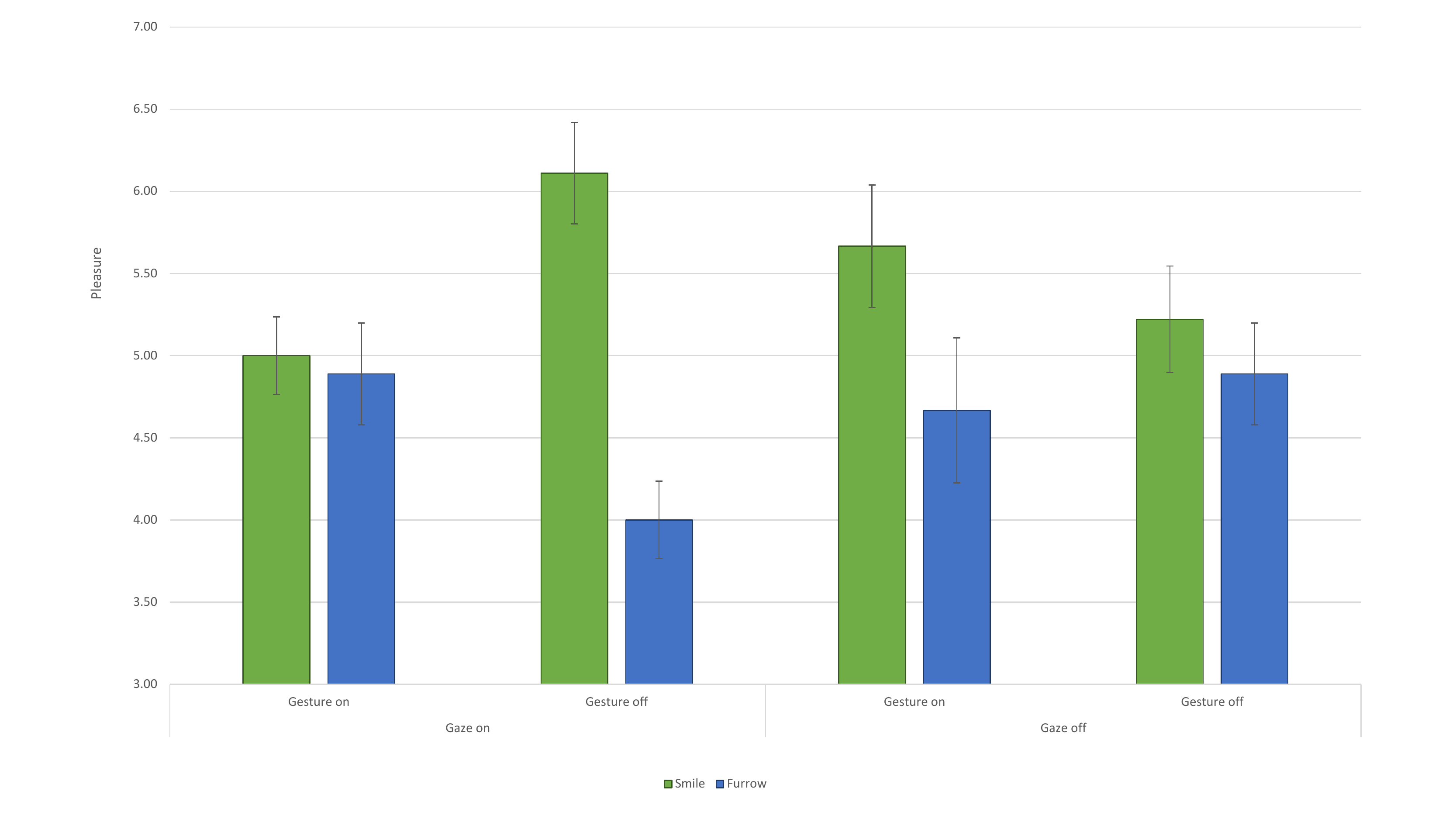}
  \caption{Three-way gaze x gesture x face interaction for pleasure ratings }
\label{fig:pleasure}
\end{figure}
}

\newcommand{\figdominance}{
\begin{figure}
  \centering
 \subfloat[]{\label{fig:face}\includegraphics[width=0.24\textwidth,  trim=0in 0in 0in 0in,
  clip=true]{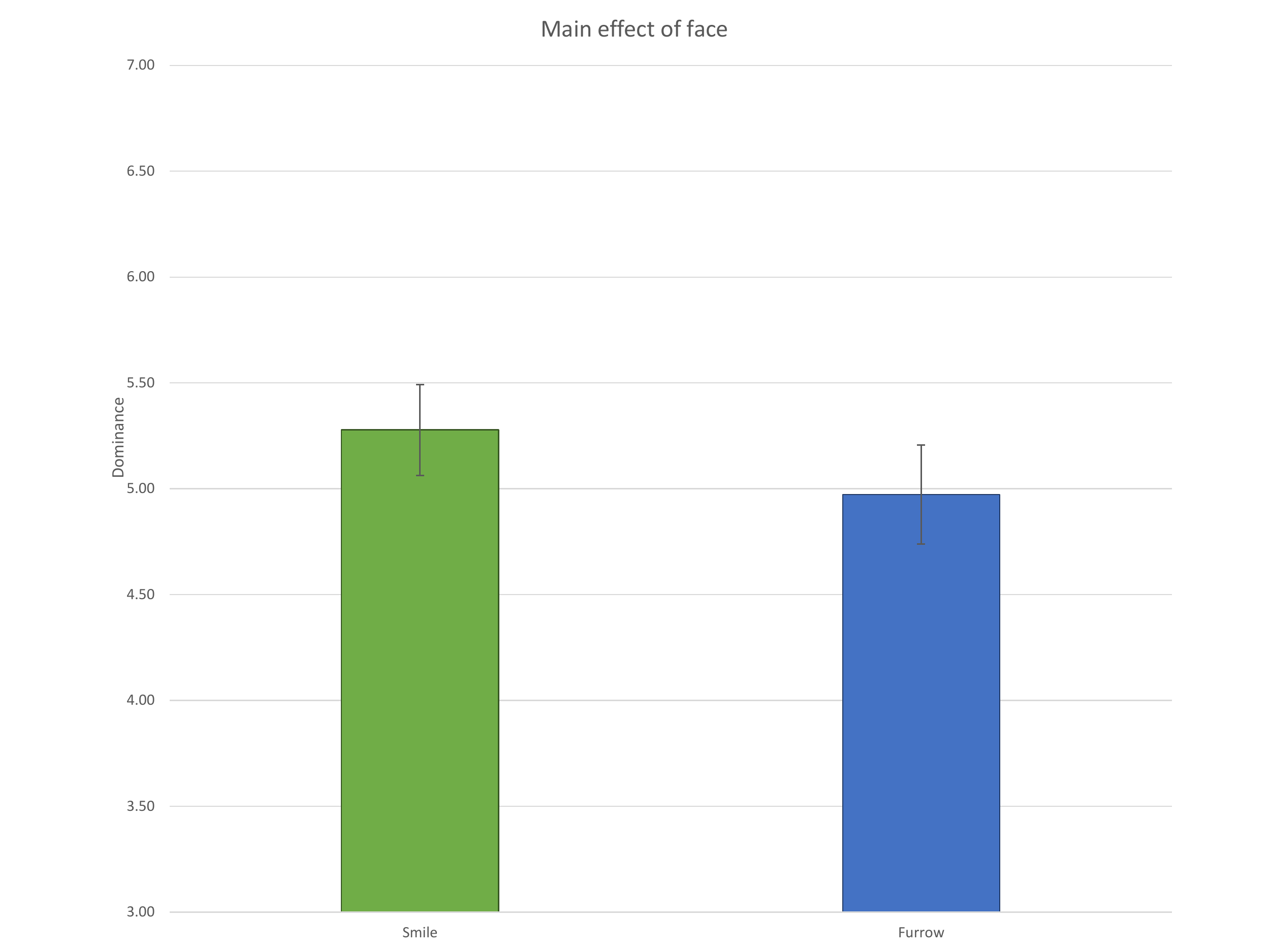}}   \subfloat[]{\label{fig:gesture}\includegraphics[width=0.24\textwidth,  trim=0in 0in 0in 0in,
  clip=true]{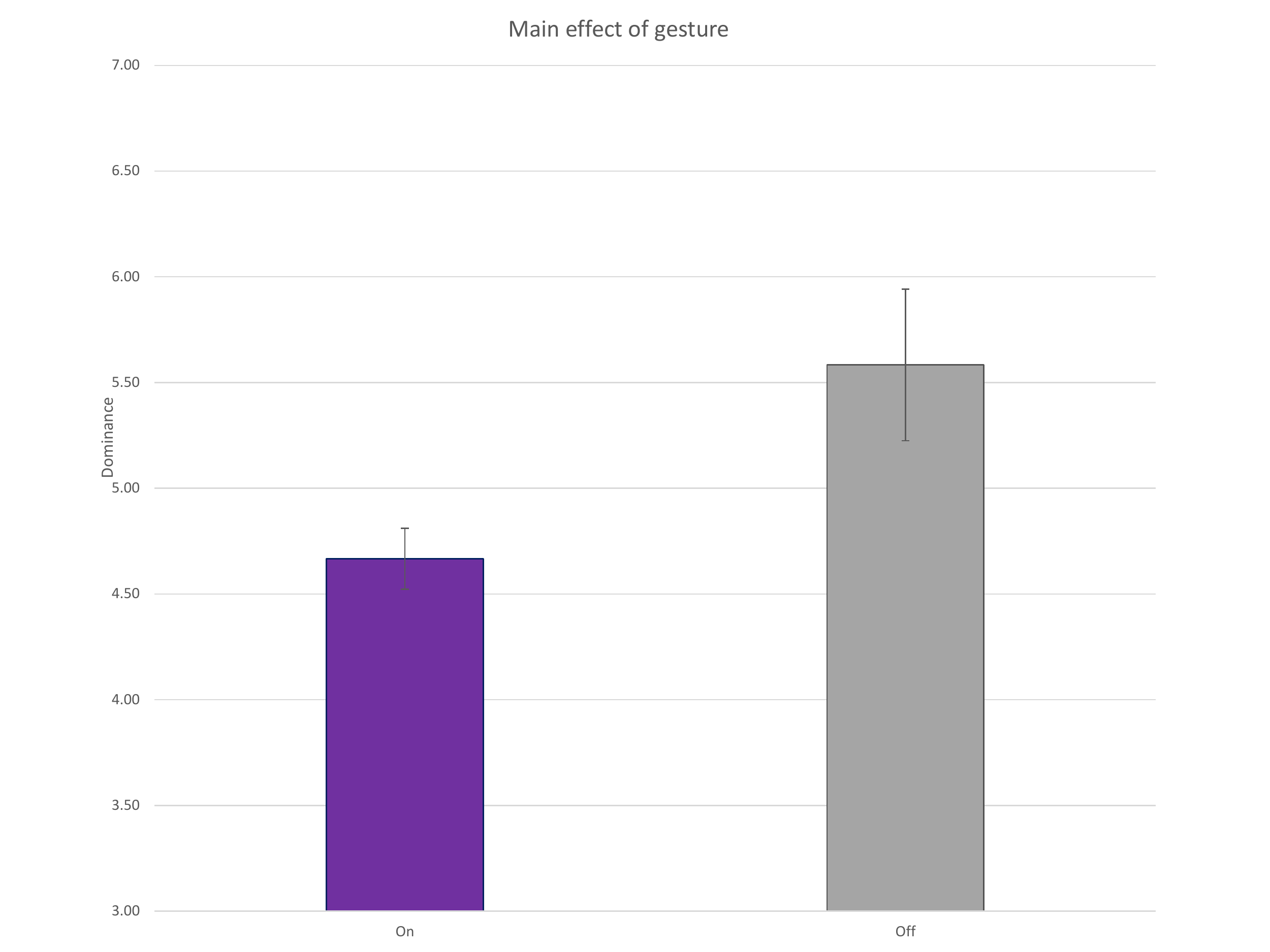}}\caption{(a) Main effect of face on dominance. (b) Main effect of gesture on dominance.}
\label{fig:dominance}
\end{figure}
}

\newcommand{\figappendixc}{
\begin{figure*}
  \centering
  \centering
    \captionsetup{labelformat=empty}
  \caption{Appendix C: Sample scenario dialogue and rating sheet}
  \includegraphics[width=0.98\textwidth,  trim=0in 0in 0in 0in, clip=true]{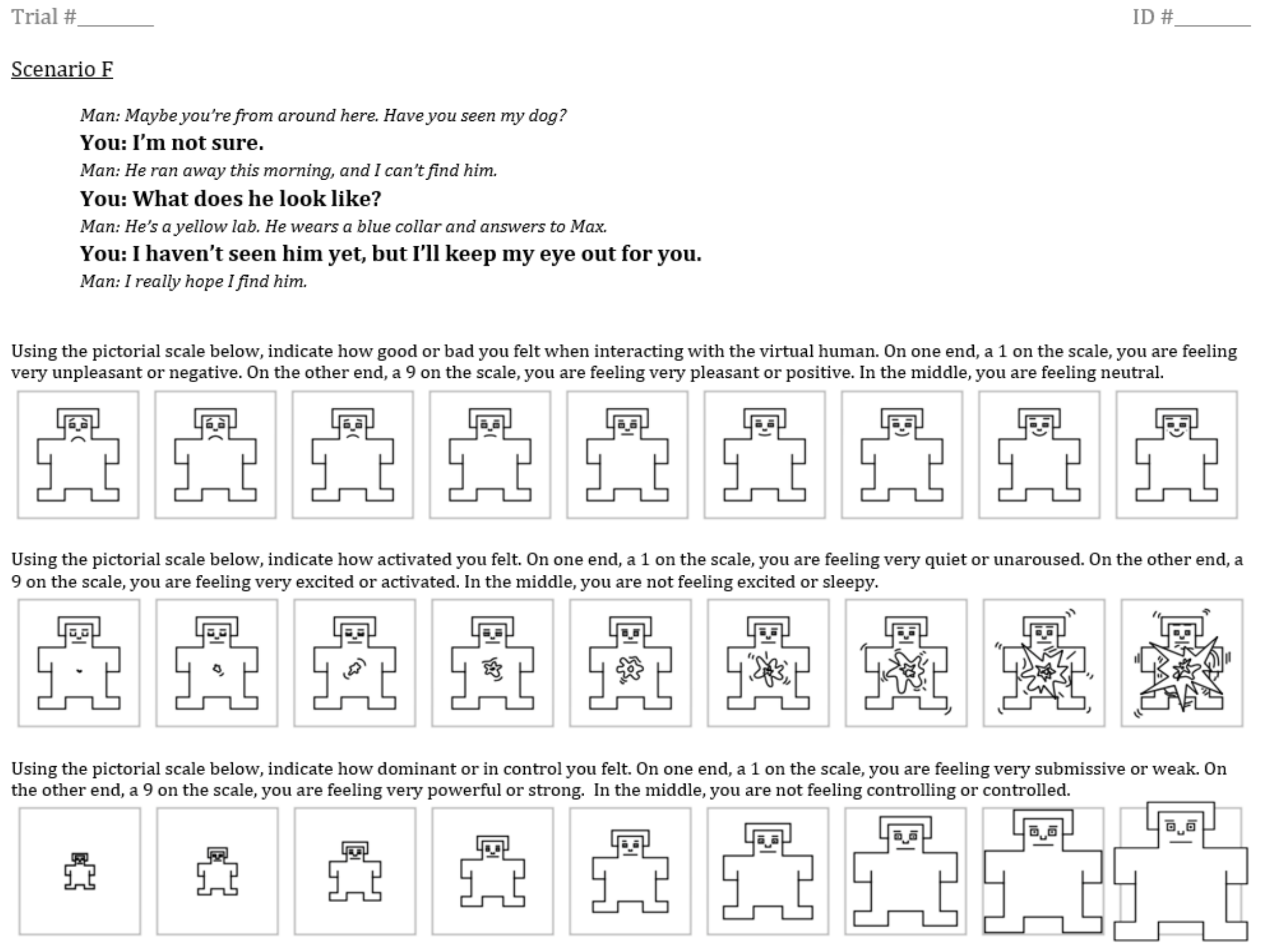}
\label{fig:appendixc}
\end{figure*}
}

\newcommand{\figappendixb}{
\begin{figure*}
  \centering
  \centering
    \captionsetup{labelformat=empty}
  \caption{Appendix B: Task instructions}
  \includegraphics[width=0.98\textwidth,  trim=0in 0in 0in 0in, clip=true]{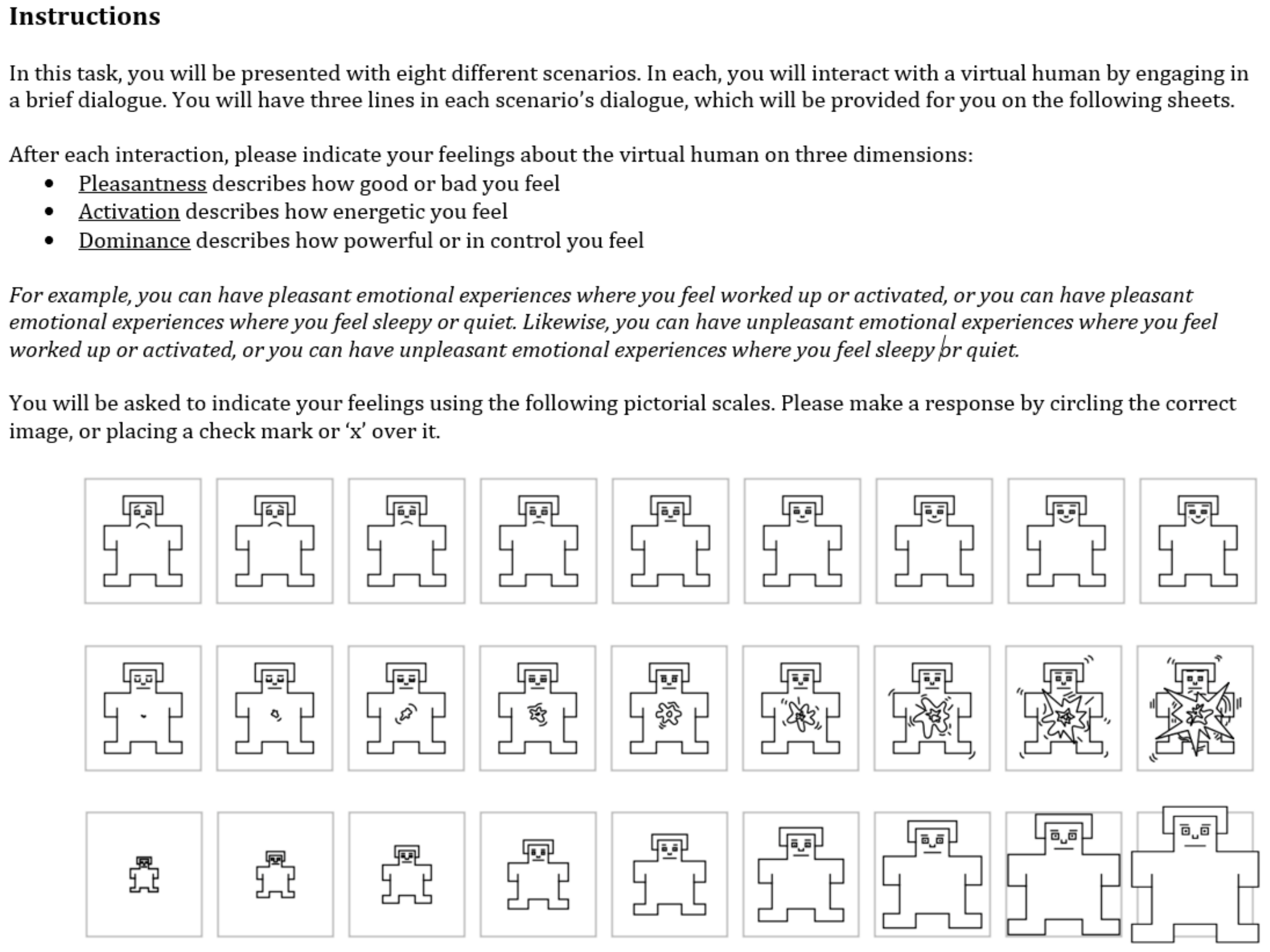}
\label{fig:appendixb}
\end{figure*}
}

\newcommand{\figappendixa}{
\begin{figure*}
  \centering
  \captionsetup{labelformat=empty}
  \caption{Appendix A: Sample dialogue}
  \includegraphics[width=0.98\textwidth,  trim=0in 0in 0in 0in, clip=true]{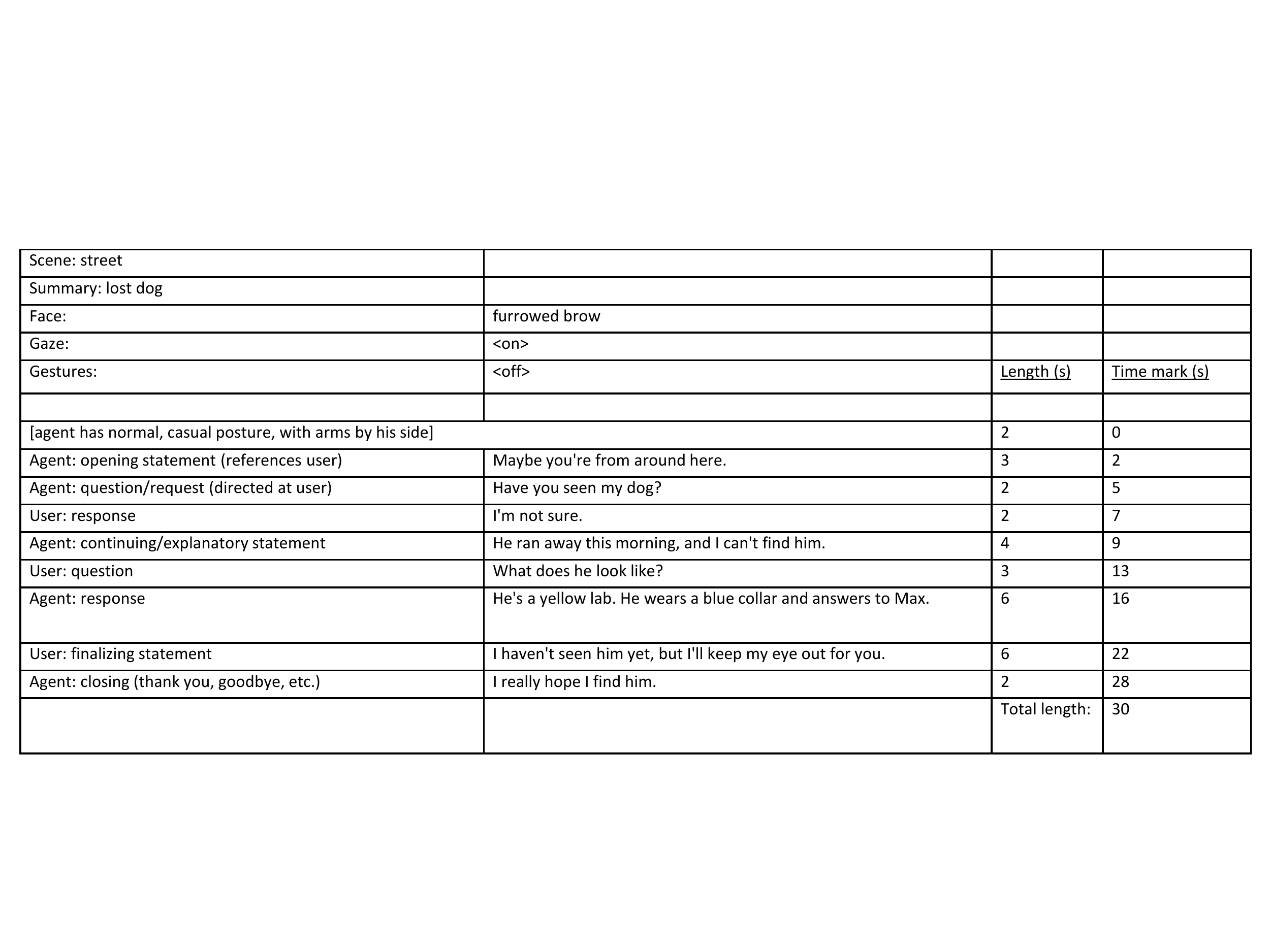}
\label{fig:appendixa}
\end{figure*}
}

\newcommand{\conditions}{
\begin{table*}
\caption{Non-verbal behaviour manipulated}
\begin{center}
 \begin{tabular}{| m{2.7cm} | m{4.85cm} | m{3.85cm} |} 
 \hline
 Condition & Level 1 & Level 2 \\
 \hline
  Facial configuration & Zygomaticus major muscle engaged\newline (‘smile’, AU 12)
 & Corrugator supercilii engaged\newline (‘furrowed brow’, AU 4) \\ 
 \hline
 Gaze pattern & Looking at participant (‘on’) & Not looking at participant (‘off’) \\ 
 \hline
 Naturalistic gesture & Present (‘on’) & Not present (‘off’) \\
 \hline

\end{tabular}

\label{tbl:conditions}
\end{center}
\end{table*}
}

\newcommand{\scenarios}{
\begin{table*}
\caption{scenarios}
\begin{center}
 \begin{tabular}{| m{1cm} | m{2.8cm} | m{1cm} |m{1cm}|m{1cm}|m{3.3cm}|} 
 \hline
 Scenario & Facial configuration & Gaze & Gesture & Scene & Dialogue \\
 \hline
 1 & Smile (AU 12) & On & On & Street & needs directions \\
 \hline
 2 & Smile (AU 12) & On & Off & House & offering tea/coffee \\ 
 \hline
 3 & Smile (AU 12) & Off & On & House & also waiting for appointment \\ 
 \hline
 4 & Smile (AU 12) & Off & Off & Street & borrowing phone \\
 \hline
 5 & Furrowed brow (AU 4) & On & On & House & showing the house \\
 \hline
 6 & Furrowed brow (AU 4) & On & Off & Street & lost dog \\
 \hline
 7 & Furrowed brow (AU 4) & Off & On & Street & asking opinion/poll \\
 \hline
 8 & Furrowed brow (AU 4) & Off & Off & House & missing agenda \\
 \hline

\end{tabular}

\label{tbl:scenarios}
\end{center}
\end{table*}
}

\newcommand{\maineffects}{
\begin{table}
\caption{Hypotheses: Main effects}
\begin{center}
 \begin{tabular}{| m{1.6cm} | m{1.3cm} | m{1.1cm} |m{1.1cm}|m{1.1cm}|} 
 \hline
 Nonverbal  & Condition & \multicolumn{3}{|c|}{Rating} \\
 \cline{3-5}
  behavior& & \underline{Pleasure} & \underline{Arousal} & \underline{Dominance} \\
 \hline
  Facial  & smile & + & - & + \\ 
  \cline{2-5}
   configuration & furrowed brow & - & + & - \\
 \hline
 Gaze pattern & on & + & - & + \\ 
 \cline{2-5}
  & off & - & + & - \\
 \hline
 Naturalistic  & on & + & - & + \\
 \cline{2-5}
 gesture & off & - & + & - \\
 \hline

\end{tabular}

\label{tbl:maineffects}
\end{center}
\end{table}
}

\newcommand{\interactions}{
\begin{table*}
\caption{Hypotheses: 2-way interactions}
\begin{center}
 \begin{tabular}{| m{2cm} | m{4cm} | m{8cm} |} 
 \hline
   & Gaze x gesture & direct gaze positive when relaxed, negative when tense  \\
 \cline{2-3}
  \underline{Pleasure} & Facial configuration x gaze & direct gaze positive when smiling, negative when furrowed brow \\
 \cline{2-3}
    & Gesture x facial configuration & gestures enhance the valence of facial configuration \\ 
 \hline
  & Gaze x gesture & averted gaze higher arousal when no gesture than with gesture \\ 
 \cline{2-3}
 \underline{Arousal} & Facial configuration x & direct gaze higher arousal when accompanied by furrowed brow \\
 \cline{2-3}
 & \textit{Gesture x facial configuration} & ? \\
 \hline
 & Gaze x gesture& greater dominance conveyed by greater kinesic involvement \\
 \cline{2-3}
 \underline{Dominance} & Facial configuration x gaze & smile is more dominant with direct gaze, furrowed brow is more dominant with averted gaze \\
 \cline{2-3}
 & \textit{Gesture x facial configuration} & ? \\
 \hline

\end{tabular}

\label{tbl:interactions}
\end{center}
\end{table*}
}

\begin{document}
%
% % paper title
% % can use linebreaks \\ within to get better formatting as desired

% make the title area
\maketitle
\thispagestyle{empty}
\pagestyle{empty}

\begin{abstract}
Human interactions are characterized by explicit as well as implicit channels of communication. While the explicit channel transmits overt messages, the implicit ones transmit hidden messages about the communicator (e.g., his/her intentions and attitudes). There is a growing consensus that providing a computer with the ability to manipulate implicit affective cues should allow for a more meaningful and natural way of studying particular non-verbal signals of human-human communications by human-computer interactions \cite{conati2002probabilistic, prendinger2005empathic}. In this pilot study, we created a non-dynamic human-computer interaction while manipulating three specific non-verbal channels of communication: gaze pattern, facial expression, and gesture. Participants rated the virtual agent on affective dimensional scales (pleasure, arousal, and dominance) while their physiological signal (electrodermal activity, EDA) was captured during the interaction. Assessment of the behavioral data revealed a significant and complex three-way interaction between gaze, gesture, and facial configuration on the dimension of pleasure, as well as a main effect of gesture on the dimension of dominance. These results suggest a complex relationship between different non-verbal cues and the social context in which they are interpreted. Qualifying considerations as well as possible next steps are further discussed in light of these exploratory findings.

\end{abstract}

\begin{keywords}
Non-Verbal Behaviour, Virtual Human, Gaze Pattern, Gesture, Facial Expression 
\end{keywords}

\IEEEpeerreviewmaketitle

\section{Introduction}
Non-verbal signals are at the heart of human-to-human interaction and communication. We nod to indicate understanding or agreement, we shift our gaze to index attention and cognitive processing, and we alter our body movements and gestures as a function of our mood or level of autonomic arousal.  The fluid production and navigation of these signals has implications for interpersonal dynamics \cite{grahe1999importance}, emotional intelligence \cite{salovey1990emotional}, and cross-cultural communication \cite{gudykunst2003cross}. Likewise, being able to naturalistically produce and follow these signals is core to creating a believable and trustworthy virtual agent \cite{blascovich2002immersive}. Producing the full range of non-verbal signals may not be desirable in all situations, however. One of the hallmarks of Autism Spectrum Disorders (ASD) is a reluctance or inability to interact socially with other humans \cite{wing1997autistic}. At the same time, individuals with ASD are known for their keen interest in technology, and readily interact with video games and non-human agents \cite{colby1971computers, milne2010development, pioggia2005android, robins2005robotic}. Research to date suggests that non-verbal signaling may be a critical factor in engaging the attention of individuals with ASD, as these signals may induce aversive physiological changes (e.g. increased autonomic arousal) to a greater extent than in neurotypical individuals \cite{georgescu2014use, goodwin2006cardiovascular, liu2008physiology}. 

While research has shown that non-human interaction partners such as virtual agents are more approachable for individuals with ASD, the exact properties that make them so have yet to be identified. The boundary conditions for ‘human-like’ interaction are critical: what behaviors or characteristics tip an agent over the line, making it off-putting? What specific aspects of non-verbal signals are being evaluated, and how does changing their presence or intensity modulate the evaluation? The ultimate goal is to find a balance of behaviors or characteristics with which the individual with ASD feels comfortable, while also working toward maximally human-like social exposure. In this manner, virtual agents can serve as pathways for teaching social interaction skills such as emotion recognition and face-to-face communication. Should non-verbal cues prove to be socio-affective roadblocks for individuals with ASD, the specific behaviors involved may feasibly also be used diagnostically, to identify levels or subclasses of the autism spectrum. If coupled with measurement of peripheral physiology, such a social-interaction-based training model could support an idiographic approach to broad-spectrum diagnosis and intervention. 

The present research seeks to address these needs by employing virtual reality (VR) or virtual agent systems to first assess neurotypical individuals’ semantic evaluations of and physiological reactions to non-verbal behaviors. As a pilot study, this project looks to gain insight into the nuances of non-verbal signaling, in order to ultimately develop a model of real-time, dynamic social evaluation that can serve as a baseline or point of comparison for individuals on the spectrum. This work extends previous social psychology and affective computing research on interpersonal dynamics and social\/affective semantic evaluation to capitalize on several advantages of VR environments. VR environments allow for tight manipulation of agent non-verbal signals while preserving immediacy and embodiment – both necessary elements to understanding the precise behavioral underpinnings and psychophysiological correlates of social interaction. Moreover, the use of virtual agents has the aforementioned advantage: the results of this investigation can be directly applied to interventions for individuals with ASD.

\subsection{Related Work}
Researchers have been doing work on how we make meaning out of social interactions and our environment for the better part of the past century.  In particular, early work by Osgood focused on what he termed the “semantic differential”: a structured data-gathering procedure in which participants rate concepts on a variety of concrete and abstract verbal scales (e.g., ‘wet vs. dry’ or ‘good vs. bad’); these ratings are then submitted to a factor analysis to determine what dimensions of meaning underlie the various judgments \cite{osgood1952nature, osgood1955factor}. Across concept categories and cultures, Osgood and his colleagues found that the same three affective dimensions – Evaluation, Potency, and Activity – consistently accounted for the variance in semantic judgments (though the exact loading of differential scales onto factors, and the variance accounted for by each dimension did shift) \cite{osgood1964semantic, tanaka1963cross}. Using similar methods, Mehrabian (later joined by Russell) interpreted these affective dimensions as Pleasure, Arousal, and Dominance \cite{russell1977evidence}. In particular, Mehrabian’s work also covered the connection between specific non-verbal behaviors and how they are evaluated along these dimensions \cite{mehrabian1968relationship, mehrabian1972measure}. In 1994, Bradley and Lang introduced the Self-Assessment Manikin (SAM), pictorial scales for quickly and intuitively capturing ratings on the dimensions of Pleasure, Arousal, and Dominance \cite{bradley1994measuring}. 

This type of dimensional model for rating stimuli has been applied widely, including in research on non-verbal behaviors and emotion perception. Critically, however, most studies to date have been based on human-to-human interaction. The use of humans as targets for socio-affective evaluation constrains experimental manipulations on what non-verbal behaviors can be naturalistically controlled, and limits the conclusions we can draw about what constitutes human-like interaction. Although work in social psychology extends the methodological toolkit by incorporating virtual environments, agents, and avatars \cite{blascovich2002immersive}, it has yet to fully explore the impact of these VR representations on participants’ affective ratings of and reactions to their social environment. For example, Bailenson and colleagues used a shared virtual environment to study the effect of head movements (as a simple non-verbal cues) on individual task performance \cite{bailenson2002gaze}. However, they failed to find a significant effect of this non-verbal channel on task performance, due to a floor effect: the task was too simple to demonstrate any improvements that may result from the knowledge of other interactants' head orientation. Work in affective computing has explored the impact of virtual agents' non-verbal signals on participant affect (e.g., \cite{bee2009relations}, \cite{cig2010realistic}, \cite{lance2007emotionally}, \cite{lance2008relation}). With the goal of developing naturalistic agent behaviors, this work often focuses on a single non-verbal channel (e.g., gaze) or single affective dimension (e.g., dominance), rather than investigating how a suite of behaviors functions when an agent is embedded in an interactive context. Similarly, although work on ASD uses virtual agents to connect with individuals \cite{milne2010development}, it has yet to investigate the precise non-verbal aspects that facilitate this communication. It is at this nexus of lines of inquiry that the current project proposal takes form.

The secondary goal of this project is to investigate the usability of physiological modalities to improve affective computing methods. Embodiment theory states that the body is an integral factor is shaping the interpretation of and reaction to the environment \cite{barrett2011beyond}, \cite{clark2008supersizing}. Likewise, theories of emotion such as psychological constructivism contend that the brain interprets the current physiological state through the process of interoception; this information is then used to predict the current emotional state \cite{barrett2014conceptual}, \cite{barrett2008embodiment}. As mentioned above, this is particularly relevant for work on ASD, as one hypothesis posits that individuals on the spectrum may experience greater autonomic arousal in social interactions, which they in turn may experience as negative affect, and so they avoid these situations \cite{georgescu2014use, goodwin2006cardiovascular, liu2008physiology}. As autonomic arousal is indexed by physiological signals such as skin conductance, respiration rate, and cardiac variables, it can be measured using standard psychophysiological methods. Psychophysiology is the practice of measuring physiological signals from the body and inferring psychological, or mental, states from them \cite{cacioppo1990inferring}. One way to make this inference is to use machine learning algorithms to produce classifiers which detect patterns in physiological responses; these patterns can then, ideally, be associated with patterns in behavioral responses (e.g., self-report) to determine their validity. To date, machine based approaches to social emotional processing have focused on facial expression and/or speech. However, we believe that physiological signals have several advantages when compared to video and sound:
\begin{itemize}
  \item The sensors used to record those signals are generally placed directly on the user, reducing potential sources of noise and problems due to the unavailability of the signal (e.g., the user not turning their head in front of the camera or not speaking);
  \item They have very good time responses: for instance, muscle activity can be detected earlier by using electromyography (EMG) than by using a camera;
  \item They cannot be easily manipulated by the user, thus minimizing the presence of ‘faked’ emotion signals;
  \item In the case of impaired users that cannot move facial muscles or express themselves, many physiological signals, such as brain waves, are still usable for emotion assessment.
\end{itemize}

In addition, there is good evidence that the physiological activity associated with affective states can be differentiated and systematically organized \cite{bradley2000emotion}. Cardiovascular and electromyogram activity have been used to examine the dimension of pleasure, or valence (i.e, positive and negative affect) of human subjects \cite{papillo1990principles, cacioppo2003social}. Electrodermal activity (EDA) has been shown to be associated with task engagement \cite{pecchinenda1996affective}. The variation of peripheral temperature caused by emotional stimuli was studied by \cite{kataoka1998development}. In this work, we exploited the dependence of one physiological response - EDA - on the underlying affective state of arousal.

\section{Approach/Methodology}
In this pilot study, we seek to develop a methodological framework for evaluating agent non-verbal behavior in a virtual reality environment. Using a semantic differential approach to gather participants’ affective ratings, and while continuously gathering physiological (EDA) data, we manipulate agent non-verbal signals in a series of short, scripted interactions.  

\subsection{Participants}
Participants were 10 university students recruited on volunteer-only basis from the Affective Computing course, as well as the investigators’ respective labs. Participants had normal or corrected-to-normal hearing and vision. We were able to balance participants by gender (5 female versus 5 male), but were not be able to control for culture or native language (4 Iranian, 3 American, 2 Indian, 1 Belgian). Mean age of the participants was 28.5 years (SD = 3.83).

\subsection{Stimuli}
Stimuli consisted of 8 short scenarios (about 30 seconds each) featuring a virtual agent who engages participants’ attention and responds to non-dynamic interactive feedback. For example, the agent may look at the participant while uttering a greeting and request (e.g., “You must be here for the open house. Can I tell you more about this room?”). Following a standard template, these scripted dialogues were written to be as neutral as possible in tone and content, and to be one-on-one interactions one might plausibly have with a (presumed) stranger. Dialogue content further corresponded with the setting in which the interaction occurred: either an indoor house/office scene, or an outdoor city street scene (see configuration details below). Although the initial plan was not to have any interaction between the participant and the virtual agent, we decided to incorporate this element in order to ensure continuous engagement of the participant, and long enough trials to gather EDA signal of adequate length.

Scenarios were developed to manipulate various aspects of the agent’s non-verbal behavior: facial configuration (smile, AU 4, or furrowed brow, AU 12); gaze pattern (looking at the participant, or looking away); and naturalistic gesture (nods and  hand movements present or absent). These 3 conditions, with 2 levels each, were controlled to produce a maximum of 8 possible combinations (see \tblref{conditions}). Scenarios are presented in \tblref{scenarios}. A random permutation of scenarios was targeted for the initial round of testing.

Stimulus generation was accomplished using the Unity game engine accompanied by Smartbody software. During design, we made use of existing assets created by the USC Institute for Creative Technologies (ICT) as a part of the VHtoolkit framework (the virtual agent, Brad, as well as the house scene), and a free asset in the Unity asset store (the street scene). Utterances to be spoken by the agent were recorded by a male speaker of American English; their corresponding lip-syncing files were generated separately and assigned to the Brad character game object. In order to control the duration of each scenario for the temporal segmentation of the physiological signals, as well as to prevent problems with the VHtoolkit-embedded speech recognizer not recognizing the participant's voices (especially non-native ones), we created a non-dynamic interaction by scripting the agent's speech at predetermined time marks. Namely, we specified dead time in proportion to the participant's utterances to allow him/her interact with the agent during each scenario. A sample dialogue, structured according to the general template and scripted by utterance time-mark, is provided in Appendix A.

For all the intended facial configurations and gestures, specific XML command executable by the SmartBody software were created and sent to the the SmartBody through the main function attached to the agent. The main function was scripted in C\# language and was used to control all the behaviours of the agent (Brad).

\figref{vignet} indicates a snapshot of one of the scenarios which take places in the house scene, the agent communicates with the participant as he smiles and looks directly toward the participant with natural gesturing.

\figvignet
\conditions
\scenarios
\subsection{Procedure}
Prior to the start of the task, participants were directed to sit at a desk, on which there is a computer monitor for presentation of the visual stimuli. Participants completed a brief demographic survey, and were then outfitted with a headset and a wristband on the non-dominant hand for peripheral physiology measurement. After establishing baseline readings of 2 minutes for the EDA signal, the researchers provided participants with written and verbal task instructions (see Appendix B for detailed instructions sheet).

In a within-subjects manipulation, participants were presented with the interactive VR scenarios. Scenarios were presented in a randomized order for a total of 8 trials. On each trial, participants were asked to attentively watch the vignette presented and speak the scripted answerer to the agent during the presentation. Afterward, they rated their reaction to the agent (i.e., how they felt based on their interaction with the agent) using the dimensional affective rating instrument provided to them on paper. For this study, the 9-point version of the Self-Assessment Manikin (SAM) was used to capture ratings on the dimensions of pleasure, arousal, and dominance \cite{bradley1994measuring} (see Appendix C for a sample scenario dialogue and rating sheet). A between-trial resting period was included to facilitate a return to baseline for the physiological signal. The physiological activity of each participant was recorded for the duration of the task using the E4 wristband sensor from the Empatica company. 

Following stimulus presentation, participants completed a basic debriefing questionnaire, in which they provided feedback about the task and the stimuli. The total running time for the procedure was 20 to 30 minutes. Participant demographics and rating data were digitized for analysis, along with any study notes. Data from one trial (scenario D: gaze off, gesture off, face smiling) were dropped due to confusion on the part of the participant. The obtained EDA signals were then segmented in several trials (or epochs), each one being associated to a given stimulus.

\subsection{Hypotheses}
Broadly, we hypothesize that each of the agent's non-verbal behaviors will impact the affective ratings provided by participants. More specifically, we predict there will be main effects such that the two levels of every condition have the opposite effect on affect. For example: a facial configuration of smile (level 1) will result in increased pleasure, decreased arousal, and increased dominance; whereas a facial configuration of furrowed brow (level 2) will result in decreased pleasure, increased arousal, and decreased dominance. These hypothesized are based on previous work in the social psychological, communication, and affective computing literature (e.g., \cite{adams2005effects}, \cite{argyle1976gaze}, \cite{bee2009relations}, \cite{burgoon1999nonverbal}, \cite{cig2010realistic}, \cite{knutson1996facial}, \cite{lance2007emotionally}, \cite{lance2008relation}, \cite{mehrabian1968referents}, \cite{mignault2003many}). Specific per-condition by level hypotheses are summarized in \tblref{maineffects}.

We further hypothesize that there will be a series of two-way interactions between levels of non-verbal behaviors. For example, in the dimension of pleasure, we expect a combination of direct gaze and a smile to result in more positive feeling, while direct gaze and a furrowed brow would result in more negative feeling \cite{adams2005effects}. We predict that gaze will likewise magnify the main effect of naturalistic gesture, and that gestures will enhance the valence set by facial configuration, as a sign of social/affective immediacy \cite{mehrabian1968referents}. See \tblref{interactions} for descriptions of additional two-way interactions. Based on a review of the above-mentioned literature, we were unable to form hypotheses about the effect the interaction of gesture and facial interaction will have on the dimensions of arousal and dominance. The results may also reveal three-way interactions, though we do not have strong a priori hypotheses on this point. As this is an exploratory pilot, we expect the findings and feedback we receive to inform these hypotheses for future studies.

Related to peripheral physiology, we predict that participants’ affective ratings will be positive correlated with autonomic arousal levels, as inferred from EDA. To wit, the mean skin resistance over a trial has been shown to be negatively correlated with the arousal of a stimulus \cite{lang1993looking}. An aroused emotion should also induce a decrease in the Galvanic Skin Resistance (GSR) signal. The increase and decrease of heart rate (HR) (here, as measured through PPG) is associated with many emotions, however an increase in heart rate indicates an overall increase in sympathetic nervous system activity and a decrease in heart rate indicates that the parasympathetic nervous system is recovering the person to a relaxation state \cite{healey2014physiological}.

\maineffects
\interactions

\section{Evaluation/Results}
\subsection{Behavioral Data Results}
The data from this 2x2x2 within-subjects design were analyzed using repeated measures ANOVAs with 3 within-subjects factors: gaze (on vs. off), gesture (on vs. off), facial configuration (smile vs. furrowed brow). Participants’ pleasure, arousal, and dominance ratings were analyzed separately as the dependent variables. 

\underline{Pleasure}
The three-way interaction between the three factors was found to be significant at $F(1,8) = 23.273$, $p = .001$, $\eta_{p}^{2} = .744$. See \figref{pleasure} for a means plot of this interaction. Follow-up 2x2 ANOVAs were conducted for each factor broken down by level:
\begin{itemize}
    \item A 2x2 ANOVA for gaze x gesture at face = smile revealed a significant two-way interaction at $F(1,8) = 12.250$, $p = .008$, $\eta_{p}^{2} = .605$. 
    \begin{itemize}
        \item Follow-up pairwise t-tests revealed that when gaze is on and the face is smiling, participants rate their experience as more pleasant when the agent gesture is off ($M = 6.100$,$ SD = .876$), compared to when it’s on ($M = 4.800$, $SD = .919$): $t(9) = -4.333$, $p = .002$. In contrast, when gaze is off and the face is smiling, the effect of gesture is not significant ($p > .05$).
        \item When gesture is off and face is smiling, participants rate their experience as more pleasant when the gaze is on ($M = 6.111$, $SD = .928$), compared to when it’s off ($M = 5.222$, $SD = .972$): $t(8) = 2.530$, $p = .035$. When the gesture is on and the face is smiling, there is no effect of gaze ($p > .05$).
    \end{itemize}
    \item A 2x2 ANOVA for gaze x gesture at face = furrowed brow revealed a significant 2-way interaction at $F(1,8) = 11.172$, $p = .009$, $\eta_{p}^2 = .554$.
    \begin{itemize}
        \item Follow-up pairwise t-tests revealed that when gaze is on and the face has a furrowed brow, participants rate their experience as more pleasant when the agent gesture is on ($M = 5.000$, $SD = .943$), compared to when it’s off ($M = 4.100$, $SD = .738$): $t(9) = 3.857$, $p = .004$.  When the gaze is off and the brow is furrowed, there is no effect of gesture ($p > .05$).
        \item When the brow is furrowed, regardless of the gesture is on or off, there is no effect of gaze (all $p^{’}s > .05$). 
    \end{itemize}
    \item A 2x2 ANOVA for gaze by face at gesture = on did not reveal a significant interaction ($p > .05$).
    \item A 2x2 ANOVA for gaze by face at gesture = off revealed a significant 2-way interaction at $F(1,8) = 11.636$, $p = .009$, $\eta_{p}^{2} = .593$.
    \begin{itemize}
        \item Follow-up pairwise t-tests revealed that when gaze is on and gesture is off, participants rate their experience as more pleasant when the agent is smiling ($M = 6.100$, $SD = .876$) than when he has a furrowed brow ($M = 4.100$, $SD = .738$): $t(9) = 5.071$, $p = .001$. When gaze and gesture are both off, there is no significant effect of face ($p > .05$).
        \item When gesture is off and the face is smiling, participants rate their experience as more pleasant when the gaze is on ($M = 6.111$, $SD = .928$), compared to when it’s off ($M = 5.222$, $SD = .972$): $t(8) = 2.530$, $p = .035$. When gesture is off and the face has a furrowed brow, there is no significant effect of gaze ($p > .05$).
    \end{itemize}
    \item A 2x2 ANOVA for gesture by face at gaze = on revealed a significant 2-way interaction at $F(1,8) = 32.029$, $p < .001$, $\eta_{p}^{2} = .781$. 
    \begin{itemize}
        \item Follow-up pairwise t-tests revealed that when gaze is on and gesture is on, there is no effect of face ($p > .05$). In contrast, when gaze is on and gesture is off, participants rate their experience as more pleasant when the agent is smiling ($M = 6.100$, $SD = .876$) compared to when he has a furrowed brow ($M = 4.100$, $SD = .738$): $t(9) = 5.071$, $p = .001$.
        \item When gaze is on and the face is smiling, participants rate their experience as more pleasant when gesture is off ($M = 6.100$, $SD = .876$) than when it is on ($M = 4.800$, $SD = .919$): $t(9) = -4.333$, $p = .002$. When gaze is on and the face has a furrowed brow, however, participants rate their experience as more pleasant when gesture is on ($M = 5.000$, $SD = .943$) than when it is off ($M = 4.100$, $SD = .738$): $t(9) = 3.857$, $p = .004$. 
    \end{itemize}
    \item A 2x2 ANOVA for gesture by face at gaze = off did not reveal a significant interaction ($p > .05$). 
    \end{itemize}

\figpleasure

\underline{Arousal}
A 2x2x2 ANOVA revealed no significant interactions or main effects (all $p^{’}s > .2$).

\underline{Dominance}
A 2x2x2 ANOVA revealed a significant main effect of gesture at $F(1,8) = 7.118$, $p = .028$, $\eta_{p}^{2} = .471$. Figure 3a: Participants indicated that they felt more dominant when interacting with the agent when gesture was off ($M = 5.583$, $SE = .358$) than when gesture was on ($M = 4.667$, $SE = .144$). 

A main effect of face was also found to be approaching significance at $F(1,8) = 4.566$, $p = .065$, $\eta_{p}^{2} = .363$. \figref{gesture}: Participants indicated that they felt more dominant when interacting with the agent when he was smiling ($M = 5.278$, $SE = .214$) than when he had a furrowed brow ($M = 4.972$, $SE = .234$).

\figdominance

 \subsection{Electrodermal Activity Evaluation}
 Electrodermal activity consists of two main components: tonic response and phasic response. Tonic skin conductance refers to the ongoing or the baseline level of skin conductance in the absent of any particular discrete environmental events. Phasic skin conductance refers to the event-related changes that are caused by momentary increase in skin conductance (resembling a peak superimposed on tonic skin conductance). In order to have a valid acquisition of the electrodermal activity, it is sampled at a higher frequency (more than 100 Hz), then filtered and down-sampled by a specific order. However, the E4 wristband EDA sensor that we used for this experiment has a frequency sampling of 4 Hz, which is too low to provide an accurate and reliable measurement of electrodermal activity.
 
 To achieve good classification results with pattern recognition and machine learning, the set of input features is crucial. For EDA signals, the time domain is most often employed for feature extraction. Consequently, we have chosen a range of features from time domain: mean, absolute deviation, standard deviation (SD), variance, and skewness.
To define an optimal set of features, a criterion function should be defined. However, no such criterion function was available in our case.
\subsection{Classification Results}
A wide range of methods has been used to infer affective states. Most of them are part of the machine learning and pattern recognition techniques. Classifiers like k-Nearest Neighbors (k-NN), Linear Discriminant Analysis (LDA), neural networks, Support Vector Machines (SVMs) and others \cite{bishop2006pattern, duda2001pattern} are useful to detect emotional classes of interest. Regression techniques \cite{bishop2006pattern} can also be used to obtain continuous estimation of emotions – for instance, in the valence-arousal space. Prior to inferring emotional states it is important to define some physiological features of interest. It is very challenging to find with certainty some features in physiological signals that always correlate with the affective status of users. Those variables frequently differ from one user to another and they are also very sensitive to day-to-day variations, as well as to the context of the emotion induction. To perform this selection, researchers generally apply feature selection or projection algorithms like Sequential Floating Forward Search (SFFS) or Fisher projection. In this project we trained a SVM-based classifier for estimation of the arousal dimensions of the participants according to their EDA signals in three different levels.

We have used the MATLAB environment and an SVM and Kernel methods (KM) toolbox, for experimenting with SVMs. The kernel function of SVM characterizes the shapes of possible subsets of inputs classified into one category. Usually, in order to fit a better classifier to the data set, a polynomial kernel with dimensionality d, defined as \eqnref{1}, is applied, but it gave us very low accuracy so we did not use any kernel for training the classifier.
\begin{equation} \label{eqn:1}
 K_{p} (x_{i} , x^{l}) = (x_{i} . x^{l})^{d}  
\end{equation}

 where $x_{i}$ is a feature vector that has to be classified and $x^{l}$ is a feature vector assigned to a class (i.e., the training sample). 
 
We trained the SVM classifier with 40 iterations to classify our data to three different classes for low, medium, and high arousal levels. In each iteration, we used a random subset of the recorded signals of 7 subjects as the training set, and the remaining 3 subjects as the test set. The average accuracy of the classifier was  57.2\%.
 
\section{Discussion/Future Work}
The results of the current work demonstrate that the relationship between different nonverbal cues can be a complex one. That is, the influence of each individual factor on the dimensions of pleasure, arousal, and dominance is not necessarily straightforward. In the present paradigm, the dimension of pleasure was most sensitive to influence from the agent's nonverbal behaviors. This is not surprising given the ease of assessment of pleasure or affective valence; anecdotally, participants readily rate stimuli on this dimension as opposed to others. A main effect of facial configuration on pleasure emerged from the data, such that participants found the agent more pleasant when he was smiling ($M = 5.500$, $SE = .228$) than when he had a furrowed brow ($M = 4.611$, $SE = .229$): $F(1,8) =28.247$, $p = .001$, $\eta_{p}^{2} = .779$. This effect is consistent with our hypotheses and, moreover, serves as a sort of manipulation check that participants understood the task instructions and were paying attention to the stimuli. In contrast, our hypotheses for the main effects of gaze and gesture on pleasure were not borne out by the data. Returning to our predictions of two-way interactions for pleasure, we find partial support for the gaze x facial configuration hypothesis that eye contact would be considered positive when the agent is smiling, but negative when he has a furrowed brow. Although this interaction was observed, it was only when the agent was not gesturing. Similarly, the gaze x gesture prediction that eye contact would be positive when the agent is gesturing, but negative when the agent is not, was only supported when the agent had a furrowed brow. In fact, the opposite influence of gesture was also observed: participants reported actually feeling less pleasant when the agent smiled, maintained eye contact with them, and gestured (vs. did not gesture). Taken together, these results suggest that gesture has a distracting influence on the overall pleasantness - contrary to our hypothesis that gestures would rather enhance the valence set by e.g. the facial configuration. 

In comparison with the dimension of pleasure, nonverbal behaviors' effects on the dimensions of arousal and dominance were less nuanced. Unfortunately, none of our hypotheses for the arousal dimension were borne out by the data, and no main effects of gaze, gesture, or facial configuration were observed. Given that this affective dimension was to be correlated with the EDA signal obtained during the task, these findings are particularly disappointing. They are not, however, particularly surprising: arousal manipulations are difficult to consistently induce in the lab, especially with short, similar, non-immersive stimuli. Further, participants find the dimension of arousal a less intuitive to grasp, and range of use issues are often anecdotally reported with arousal rating scales. For the dimension of dominance, we did see an expected main effect of face, and in the predicted direction: participants reported feeling more dominant when the agent was smiling than when he had a furrowed brow. This finding fits with our experience that interaction partners who are happy (as often inferred from a smile) are not as threatening as interaction partners who are angry (as often inferred from a furrowed brow). Considered from an approach\/avoid distinction, both happiness and anger are considered 'approach-motivated' emotions, and would thus engage more immediate feelings of dominance or submission within a social encounter \cite{marsh2005effects}. Lastly, although we did see a trending main effect of gesture on dominance, it was opposite that predicted. In contrast to the hypothesis that participants would perceive a lack of gesture as more threatening, they reported feeling less dominant when naturalistic gesture was present. Despite going against our predictions, it is explained nicely by the association of referential gestures and nods with physical and social immediacy \cite{mehrabian1968referents}, which could be considered a form of threat.

Overall, the present results offer null or somewhat inscrutable findings due not only to the multi-layered relationship between nonverbal behaviors, but also due to the nature of the present study. Pilot work involves a reduced sample size (here only n=10) and incompletely developed stimuli, both contributing to a small, noisy data set. As further confounds for interpretation, the current stimuli also incorporated elements other than the agent - namely, the dialogue content and the background scene - which undoubtedly contributed to participants' experience of the scenarios. For example, a conversation with a stranger about a lost dog is likely inherently more pleasant - even if the person is scowling the whole time - than a request from a pollster for your opinion. More generally, scenarios in which the agent is asking if he can do something for the participant (e.g., bring him/her a beverage) are presumably going to be more enjoyable than scenarios in which the agent needs something from the participant (e.g., borrow his/her phone). (Indeed, pairwise t-tests based on this property reveal that participants reported feeling significantly more pleasant when the agent was doing them a favor ($M = 5.533$, $SD = .632$) than when he was requesting something from them ($M = 4.705$, $SD = .671$): $t(9) = 5.459$, $p < .001$.) Moreover, the context in which an interaction occurs also influences one's interpretation. During the feedback process, in fact, one participant reported feeling more pleasant during the indoor house scenes than during the outdoor street scenes, as the indoor space was brighter and had a more protected feel (presumably because interactions inside a building are more of one's choosing). Future iterations of this work should seek to address these possible confounds, for example by randomizing the pairing of dialogue and nonverbal behavior, controlling for interaction type (favor vs. request), or separately norming background scenes for pleasure and arousal. 

Future work should also seek to address procedural and technological shortcomings of the present work. Feedback from participants indicated that they found it difficult to focus on and fully evaluate the virtual agent because they were constantly looking back to the scripted dialogue on the paper in front of them. Though this item can be addressed in the short term by having the scripts available on the computer monitor, alongside the screen with the agent, ultimately this problem would be fully resolved by incorporating dynamic speech recognition and response within the VR environment. Dynamically generated speech would also resolve issues with timing of speech (e.g., sometimes the pre-programmed dead space was too long, sometimes too short) and would allow participants to have a fully immersed, authentic conversation with the agent. The use of a VR headset would complete the immersion in the environment (although rating instruments would somehow need to be incorporated). For more accurate measurement of electrodermal activity, the E4 wristband would either be augmented with a separate set of electrodes, or replaced by wet sensors as used in laboratory recordings. For fuller description and classification of psychophysiological data, additional channels such as heart rate or blood volume pulse could be captured - again, ideally with laboratory recording equipment rather than wearable technology. All told, the current work provides a useful starting point for various possible future studies, each able to address a different aspect of the same overarching goal: to use virtual humans to better understand real ones.

\section{Acknowledgements}
\balance
\bibliographystyle{IEEEtran}
\small
Both Behnaz Rezaei and Katie Hoemann designed the procedure: Katie selected the psychological measures, and Behnaz selected the physiological measures. Both Katie and Behnaz designed and developed the virtual agent characteristics and VR scenarios. Katie wrote the scenario dialogues, recorded and cleaned the audio files, and contributed to asset configuration and BML code generation for the virtual character. Behnaz completed all coding and scripting of the scenarios, along with additional asset configuration.

Katie was the primary author for the introduction, related work, psychological approach/methods, behavioral results and discussion sections. Behnaz was the primary author for the physiological approach/methods, and physiological data evaluation/results sections.

The authors would like to thank Dr. Stacy Marsella for his continued support and suggestions for the appropriate software platform in which to realize this project. Dan Feng was instrumental in providing support throughout the development and testing process, including generation of the BML files for naturalistic speech. This project would not have succeeded - and we would not have learned nearly as much - without her involvement.

\bibliography{paper}

\newpage
%\appendix 
%Appendix A:Sample dialogue
% the \\ insures the section title is centered below the phrase: AppendixA
\figappendixa

\newpage
%\chapter{Appendix B: Task instructions}
% the \\ insures the section title is centered below the phrase: Appendix B

\figappendixb

\newpage
%Appendix C: Sample scenario dialogue and rating sheet

\figappendixc

% that's all folks
\end{document}